\documentclass{aa}
\usepackage{graphicx}
\usepackage[varg]{txfonts}
\usepackage{longtable}
\usepackage{lscape}

\begin{document}

 \title{Unexpectedly strong effect of supergranulation on the detectability of Earth twins orbiting Sun-like stars with radial velocities}

   \titlerunning{Effect of supergranulation on the detectability of Earth twins orbiting Sun-like stars}

   \author{N. Meunier \inst{1}, A.-M. Lagrange \inst{1}
          }
   \authorrunning{Meunier et al.}

   \institute{
  Univ. Grenoble Alpes, CNRS, IPAG, F-38000 Grenoble, France\\
\email{nadege.meunier@univ-grenoble-alpes.fr}
             }

\offprints{N. Meunier}

   \date{Received ; Accepted}

\abstract{Magnetic activity and surface flows at different scales pertub radial velocity measurements. This affects the detectability of low-mass exoplanets.}
{In these flows, the effect of supergranulation is not as well characterized as the other flows, and we wish to estimate its effect on the detection of Earth-like planets in the habitable zone of Sun-like stars. }
{We produced time series of radial velocities due to oscillations, granulation, and supergranulation, and estimated the detection limit for a G2 star and a period of 300 days. We also studied in detail the behavior of the power when the signal of a 1~M$_{\rm earth}$ planet was superposed on the signal from the stellar flows. }
{We find that the detection rate does not reach 100\% except for the supergranulation level we assume, which is still optimistic, and for an excellent sampling. }
{We conclude that with current knowledge, it is a very challenging task to find Earth twins around Sun-like stars with our current capabilities. }


\maketitle

\section{Introduction}

Magnetic activity produces spurious radial velocities (hereafter RV) and thus affects the detectability of exoplanets \cite[e.g.,][]{saar97,hatzes02,saar03,wright05,desort07,lagrange10b,meunier10a,meunier13,dumusque16,dumusque17}. Oscillations, granulation, and supergranulation  \cite[][]{dumusque11b,meunier15}  also lead to RV variations, with amplitudes at the 1~m/s level: these effects may hide   exoplanet light signals in RV time series. Averaging techniques have been tested to mitigate these effects. Oscillations can be relatively well averaged out \cite[][]{chaplin19}. Although it weakens the effect of granulation and supergranulation on RV \cite[][]{dumusque11b}, averaging applied for many hours in the solar case did not allow completely eliminating these signals \cite[][]{meunier15}. 


Although oscillations and granulation are well characterized for the Sun and are the most easily averaged  because of their short associated timescales, supergranulation has a longer lifetime and is potentially more problematic. The origin of supergranulation is still unclear \cite[see the reviews by][]{rieutord10,rincon18}.  Typical flows are slower than for granules, with velocities of a few 100 m/s,  and are mostly horizontal, but their spatial scale is larger, about 30 Mm \cite[e.g.,][]{hart56,leighton62}. At any given time, there are far fewer supergranules than granules, therefore the resulting RV jitter associated with supergranules remains large.  The supergranulation scale appears to be strongly related to the large-scale dynamics of granules
\cite[e.g.,][]{rieutord00,roudier16}. 
Furthermore, both granulation and supergranulation are expected to exhibit some power at low frequency, as suggested by the shape of the power spectrum proposed by \cite{harvey84}. The simulation of \cite{meunier15} confirmed this shape, and the parameters for a few stars other than the Sun have been estimated by  \cite{dumusque11b}.

We wish to estimate the effect of these flows on the detection of Earth-like planets orbiting in the habitable zone around G2 stars. After describing the model and the analysis method, we show the strong effect of supergranulation on the detectability of such long-period planets.  We also estimate the effect of this signal on the determination of the planet masses in follow-up observations of validated transits. 
Future work will extend this work to a wider range of spectral types and more complex samplings.

\section{Model}

\subsection{Time series}

We built RV time series that are produced by granulation, supergranulation, and oscillations of G2 stars as described in \cite{meunier19}: we computed the power spectrum versus frequency using well-known laws, and then the inverse Fourier transform. The resulting 30-second cadence ten-year-long time series were then binned over one hour. The parameters of the power laws are those corresponding to the Sun. Because supergranulation is less well constrained, a range of amplitudes was compatible with our current knowledge of supergranule properties according to the simulations made in \cite{meunier15}, who obtained a root-mean-square (rms) RV due to solar supergranulation between 0.28~m/s and 1.12~m/s. \cite{dumusque11b} estimated an rms of 0.77~m/s for $\alpha$ Cen A (also a G2 star), and the ratio between granulation and supergranulation power for their three main-sequence stars (spectral types between G2 and K2) is compatible with the supergranulation level between the lower estimate of \cite{meunier15} and $\sim$0.7 m/s. In this work we therefore compute the detectability for two possible extreme levels for supergranulation, which we assume represent a realistic range: the low estimate corresponding to the 0.28 m/s rms (hereafter "low"), and a level corresponding to an rms of $\sim$0.7 m/s (hereafter "med").

We considered one point per night. The full sampling therefore represents 3650 hours of observation. Then, more realistically, a gap of four months each year was introduced to simulate the fact that a given star can usually not be observed throughout the year. We considered $n$ consecutive observing nights (with $n$=1, 3, 6, 9, or 12), and randomly positioned 20 such sets of $n$ nights each year, over ten years. This leads to between 200 and 2400 hours. 

In addition to these temporal  samplings, we also simulated two other samplings, with 320 and 1500 points each, with a more regular pattern (still taking into account the four-month gap). The first sampling corresponded to a typical step between observations of about one week, while the in second sampling, the step was much shorter. A small dispersion was added to the dates, as well as a few random gaps.  

Averaging over one hour aims at mitigating granulation to a reasonable level. It does not suppress this signal completely, as we show in the next section, but it allows the rms of the RV signal to decrease from 0.83~m/s (rms for our time series with a 30 sec cadence) to 0.39~m/s. Averaging over a longer duration would allow us to decrease the jitter due to granulation further, but in a less efficient way because one hour corresponds well to the inflection point in the curve jitter versus bin size in \cite{meunier15}. 
In this condition, the remaining signal due to oscillations is naturally very low (rms below 0.03~m/s). In the case of supergranulation, the one-hour averaging is not efficient, and we show that this represents the dominant signal in these simulated data. The rms of the supergranulation signal after this one-hour averaging is still close to 0.70~m/s for medium level supergranulation and 0.28~m/s for the low level signals.


\subsection{Analysis}

\begin{figure} 
\includegraphics{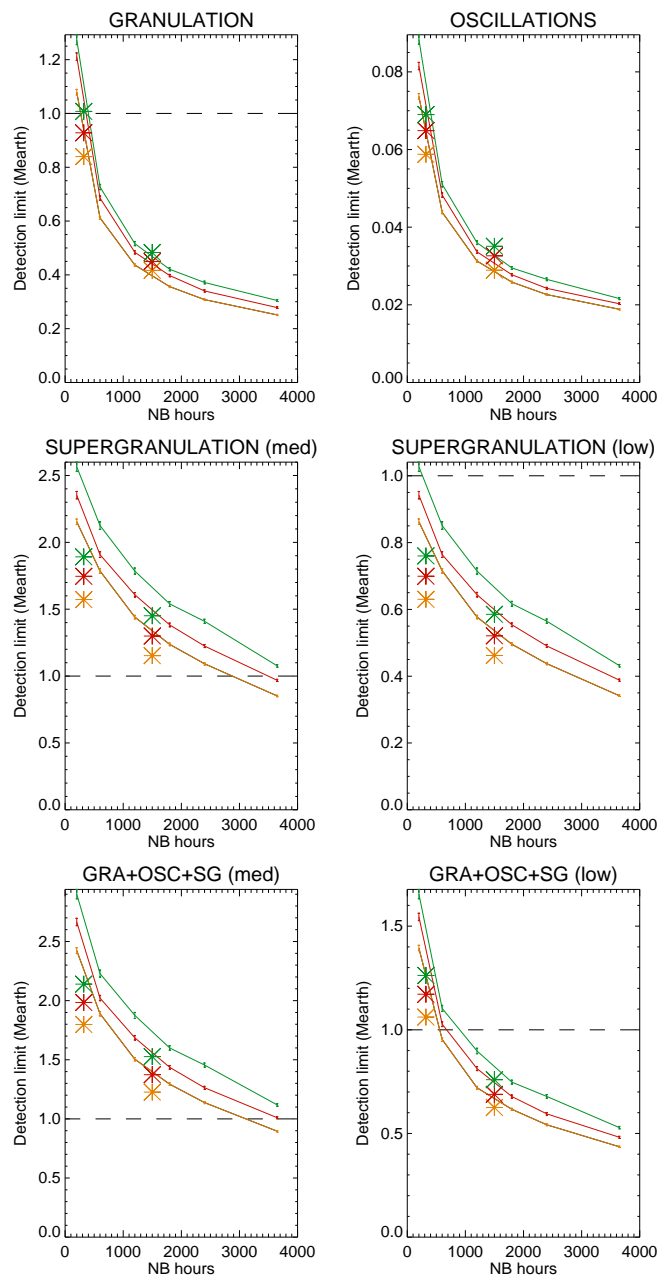}
\caption{
{\it Upper left panel}: Average LPA detection limit (averaged in phase and signal realizations, in M$_{\rm earth}$) vs. the number of hours for the inner side of the habitable zone (275d, orange), the middle (505d, red), and the far side (777d, green). The stars correspond to the additional samplings at 320 and 1500 points (same color code).
Here the planet signal is not superposed on the oscillation, granulation, and supergranulation signal (see text).  
{\it Upper right panel}: Same for the oscillation signal alone.
{\it Middle left panel}: Same for the supergranulation signal (medium level) alone.
{\it Middle right panel}: Same for the supergranulation signal (low level) alone.
{\it Middle left panel}: Same for the contribution of the three signals (medium level for supergranulation).
{\it Middle right panel}: Same for the contribution of the three signals (low level for supergranulation).
}
\label{limdet}
\end{figure}

\begin{figure} 
\includegraphics{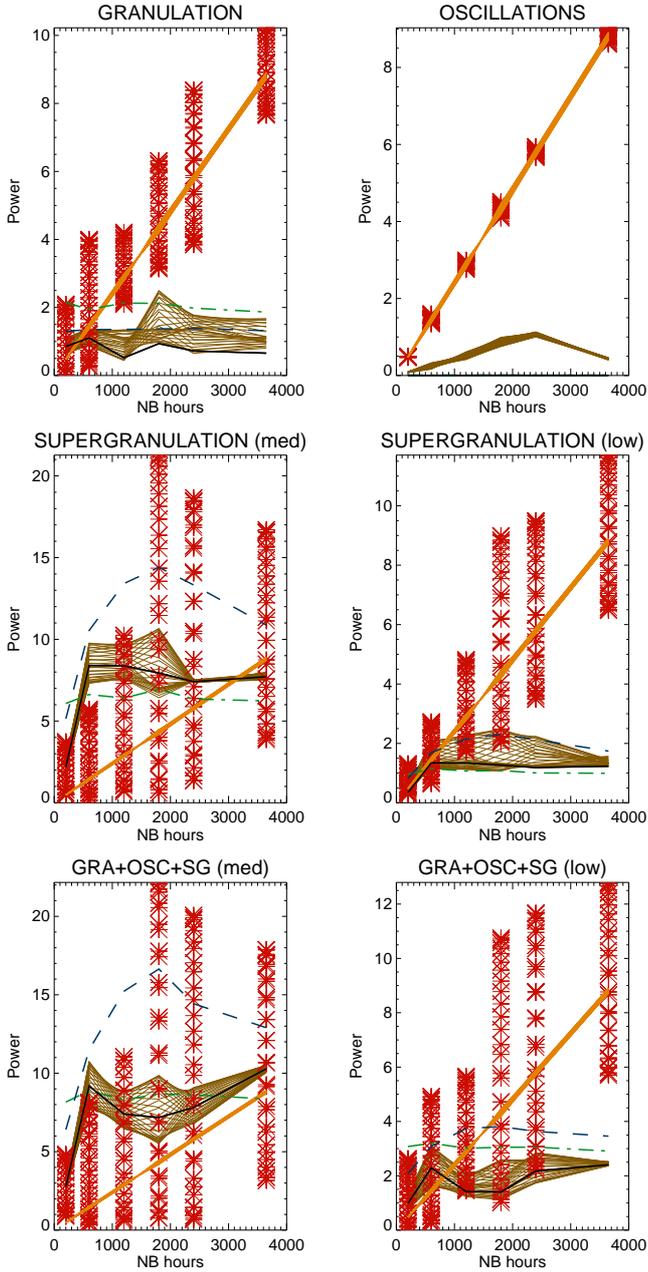}
\caption{
Power in the periodogram due to a 1 M$_{\rm earth}$ planet at 300 day (P$_{\rm pla}$) when superposed on the stellar signal (red stars). For comparison, the power due to the planet alone is shown in orange. The brown lines are the power around P$_{\rm pla}$ in the presence of the planet (for different phases), and the black lines show the power when only the stellar signal is present. The dot-dashed green line is the 1\% fap, and the blue dashed line shows the 1\% level of the stellar signal for periods longer than 100 d (P$_{\rm ogs,LF}$, see Sect.~2.2). Each panel represents a different stellar signal (similar to Fig.~1). 
}
\label{power}
\end{figure}

\begin{figure} 
\includegraphics{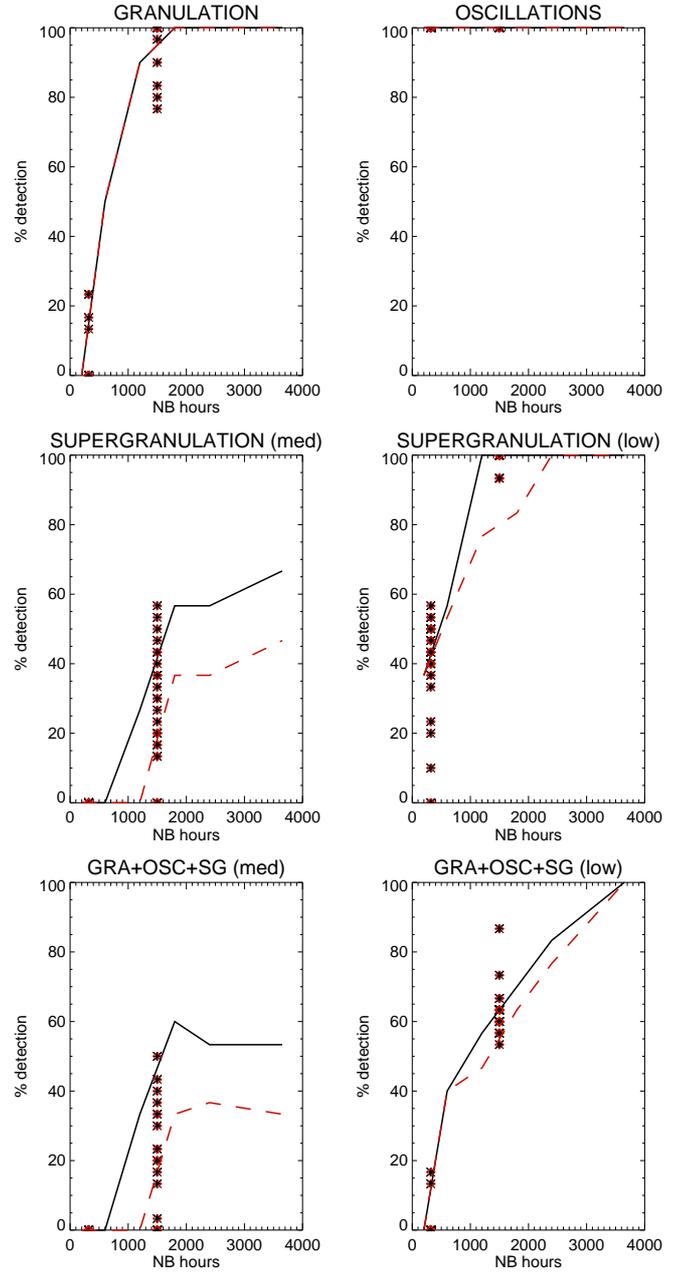}
\caption{
Percentage of detections for a 1 M$_{\rm earth}$ planet at 300 day (P$_{\rm pla}$) for two criteria: power at P$_{\rm pla}$ larger than the surrounding and larger than the fap (solid black line), and power at P$_{\rm pla}$ also larger than the stellar power P$_{\rm ogs,LF}$ (dashed red line). Black and red stars represent the percentage of detections for the 320 and 1500 point samplings for 30 realisations of the signal. In some cases, all 30 points are superposed (0\% or 100\% in particular).  
Each panel represents a different stellar signal (similar to Fig.~1). 
}
\label{pc}
\end{figure}

We computed the detection limits using the local power analysis (LPA) method that we developed in \cite{meunier12}, with examples in \cite{lannier17}. This method compares the amplitude of the power due to a planet at orbital period P$_{\rm pla}$ (in our case, assumed to be on a circular orbit, with various phases) with the maximum power in the range 0.75-1.24P$_{\rm pla}$. The detection limit is the mass for which the planet power is 1.3 times the maximum power observed in the periodogram of the stellar signal. We selected here three periods corresponding to the inner side of the habitable zone (275d), the middle (505d), and the far side (777d) for a G2 star \cite[][]{kasting93,zaninetti08,jones06}, as in \cite{meunier18c}.


This approach does not take into account the fact than when the planet signal is superposed on the stellar RV, the amplitude of the peak due to the planet is very sensitive to the phase of the planet. When it is close to the detection limit, it can be strongly enhanced or it can disappear. To quantify this effect, we proceeded as follows: we added a planetary signal (1 M$_{\rm earth}$, and P$_{\rm pla}$ of 300 days) to the stellar signal for 30 different phases. For each of these realizations, we computed from the periodogram 1/ the power at the orbital period P$_{\rm pla}$ , and 2/ the maximum power in the ranges 0.3-0.9P$_{\rm pla}$ and 1.1-3P$_{\rm pla}$, that is, just below and above the planet period. A necessary (but not sufficient) condition for a detection and characterization is that the planet-induced peak is higher than the power surrounding it, otherwise the planet period cannot be characterized.

A classical method to determine the significance of a given peak is to compute a false-alarm probability (fap) using a bootstrap analysis and then the maximum power over the considered range in period. For a detection, the power at the planet period must be larger than the fap, for example, at 1\%. We computed these faps for each sampling. We performed 1000 realizations of the bootstrap.  

Finally, the main obstacle to the detection of a planet being the presence of peaks due to the stellar signal, the fap provides an indication on the possible presence of a planet only if the stellar signal behaves as white noise (assumption made during the bootstrap analysis). It does not take into account the significant peaks that could be present due to the frequency distribution of the stellar signal. The shape of the observed power spectrum of granulation and supergranulation is such that in principle, some power at long periods could be present. To quantify this effect, we computed 1000 realizations of the time series  due to oscillations, granulation, or supergranulation (using the method described in Sect.~2.1), and then the maximum power for each of these realizations over the whole range in frequency and for periods longer than 100 days only.  This gives an idea of the maximum power at long periods that can be produced by granulation or supergranulation for a large number of realizations. As for the fap, it is then possible to determine a 1\% level associated to the presence of the stellar signal for each of these 1000 values for the whole period range and above 100 days.  We call this  P$_{\rm ogs,all}$ and P$_{\rm ogs,LF}$ in the following.

\section{Results}

\subsection{LPA detection limits}

Fig.~\ref{limdet} shows the detection limit versus the number of observing hours (spread over ten years) for three planet periods covering the habitable zone of a G2 star computed for oscillations, granulation and supergranulation, either alone or added. 
For each sampling, the detection limits were computed for 1000 realizations and then averaged.
The detection limits are below 1 M$_{\rm earth}$ when granulation alone is considered (if more than 200-300 hours have been acquired) or oscillations alone are studied for all samplings. In contrast, the medium level of supergranulation leads to detection limits higher than 1 M$_{\rm earth}$ for most samplings (except for all points for which it is above 1 M$_{\rm earth}$). The lowest level alone allows us to reach detection limits lower than 1 M$_{\rm earth}$ in most cases.
When the sum of the three stellar signals is considered, the detection limit is on average lower than 1 M$_{\rm earth}$ when the number of observations exceeds 500-1000 hours for the low level of supergranulation. It is mostly above 1 M$_{\rm earth}$ for the medium level of supergranulation, however. The more regular sampling (stars in Fig.~\ref{limdet}, corresponding to 320 and 1500 points) tends to show slightly lower detection limits in some cases, but the effect is weak. Detection limits are slightly higher for the far side of the habitable zone, that is, it is dominated by the amplitude of the planetary signal itself.

\subsection{Effect of the planet phase}

The detection limits above do not take into account the strong variability of the power at the planet period with phase when the planet RV is superposed on the stellar signal (see Sect 2.2).  Fig.~\ref{power} illustrates this effect for a 1 M$_{\rm earth}$ planet at a period of 300 days for one realization of the signal and sampling: the red stars show the wide range in power that is covered at the planet period for different values of the phase. The comparison of these red stars with the fap level agrees well with the previous results; the red stars also allow us to compute the percentage of planets that can be detected for the various samplings (next section). 

We also obtain important results from these plots. 
For granulation, the green (fap) is slightly above the blue curve (P$_{\rm ogs}$, computed at long period only).
This means that the power at long periods (including at a few days) due to granulation is not very different from the power that would be produced by white noise of the same dispersion. The shape of the granulation power spectrum is therefore not critical, and the fap is a reliable criterion for determining the significance level of an observed peak that might be due to a planet. Simply decreasing the rms (by averaging or other methods) will directly improve the detectability, and the fap as well.

On the other hand, the two curves are different for supergranulation, especially for the medium level:
supergranulation produces peaks above the fap at long periods. As a consequence, it is possible to observe a peak or peaks above the fap that cannot be attributed to a planet with a reasonable level of certainty,  because the probability is high that these peaks are be due to supergranulation. It is more reliable to use P$_{\rm ogs,LF}$ to decide whether a planet could be present. When it is computed over all periods (as for the fap),  P$_{\rm ogs,all}$ is also larger than the fap for supergranulation (and for the total signal). 
 

Finally, Fig.~\ref{power} also shows that the power at periods  surrounding the planet period can also be large and above the fap, in particular when the sampling is good. 
This shows that it is possible to observe a power above the fap at periods different from the period where the power is injected, covering a wide range of periods. This makes interpreting the presence of the significant power even more difficult.

\subsection{Percentage of detection}

Fig.~\ref{pc} shows the percentage of cases for which the power at the planet period (for a 1 M$_{\rm earth}$, P$_{\rm pla}$ of 300 day) is larger than the power in the surroundings and larger than the fap (solid black line). We  assumed that if the power due to the planet is below one of these thresholds, then one would not claim the detection of such a planet. Furthermore, given the presence of a significant power for supergranulation, that is, above the fap, two previous conditions may not be sufficient to provide a reliable detection rate because peaks due to supergranulation could also correspond to such conditions. 

We therefore propose to add a third  constraint:  the power at the planet period should also be larger than the power P$_{\rm ogs,LF}$ (here at the 1\% level) due to the stellar signal (dashed blue lines in Fig.~\ref{power}). The corresponding detection rates are indicated as red lines. For granulation, a 100\% detection level is reached for more than 1200 hours of observations, and a few hundred points are sufficient to reach 50\% of success in the detection. 
For supergranulation and even more for the total signal, it is difficult to reach a rate of 100\%, however. For the medium level of supergranulation and the strongest criteria, the percentage is always lower than 60\%, even for a sampling as good as 3650 points over ten years. The lower level of supergranulation allows us to reach 100\%, but only for that excellent sampling, otherwise the success rate is much lower; it is typically below 50\% for fewer than 1000-1500 hours (depending on the criteria). 
For the two additional temporal samplings, we performed very many realizations, so that Fig.~\ref{pc} shows the dispersion in detection rates for each number of points. The sampling corresponding to the solid and dashed line is compatible with the detection rates obtained with this more regular sampling.
 
\subsection{Mass characterization}

\begin{figure} 
\includegraphics{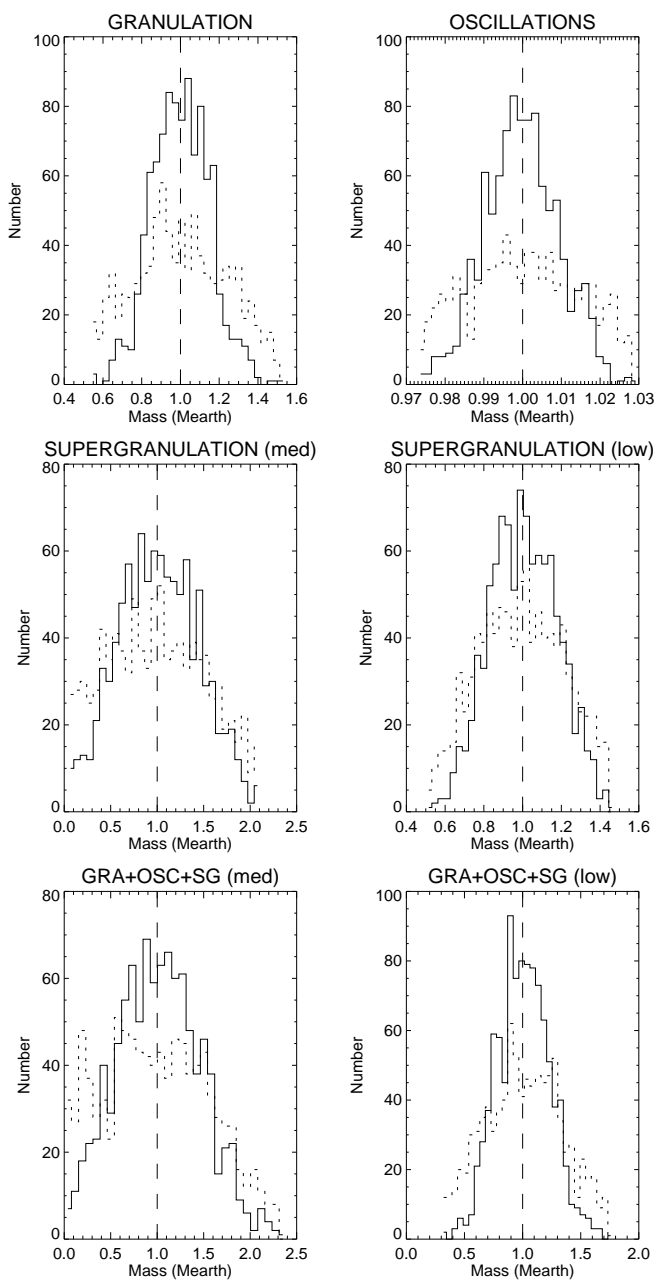}
\caption{
Mass distribution for the 1500-point sampling (solid line) and 320-point sampling (dashed line) line when the period and phase of the planet are well constrained. The vertical dashed line shows the true planet mass (1 M$_{\rm Earth}$).  
Each panel represents a different stellar signal (similar to Fig.~1).
}
\label{mass}
\end{figure}

When the RV observations are dedicated to the follow-up of well-established transits, the objective is to characterize the mass of the planet rather than to validate a detection. The problem is then different from comparing the observed amplitude with a false-positive level, as done in the previous section. Here we consider that the planet period and phase are perfectly constrained from the transit observations. For 1000 realizations of the signal and the two specific samplings at 320 and 1500 points, we fit the planetary signal (no eccentricity, 300d period) and derived the mass of the planet. The mass distributions are shown in Fig.~\ref{mass}. The characterization is improved for the 1500-point sampling compared to the 320-point sampling. For granulation alone, the rms is about 31\% and 15\% for 320 and 1500 points, respectively, the latter representing a good mass characterization. The oscillations never prevent reaching very good precisions on the mass (1-2\%). The medium level of supergranulation alone leads to uncertainties higher than those due to granulation (58\% and 43\%), while the low level is better (24\% and 17\%). When the three signal categories are superposed, the percentages are between 45\% and 62\% (for the medium supergranulation level) and between 23\% and 39\% (for the low supergranulation level).

\section{Conclusion}

The simulations for a G2 star shows that  although the presence of granulation can be mitigated if the number of observations is large but reasonable ($\sim$ 500-1000 hours), this is not so for supergranulation. 
Considered alone, our medium estimate leads to a 40-60\% detection rate for a 1 M$_{\rm earth}$ planet with a 300-day period, even for 3600 h spread over ten years. 100\% is reached only after 1200-2400 h for the low level. 
Only the optimistic estimate for the supergranulation level (rms of 0.28 m/s) for the one-hour binning and a very long observing time (ten years, one hour per night, every night) allows reaching a detection rate of 100\% for a 1 M$_{\rm earth}$ with a 300 d period (i.e., in the habitable zone) when it is also superposed on oscillation and granulation. In particular, for 1200 h (100\% rate for granulation), the detection rate for the sum of the three contributions is only in the 40-60\% range. 

Furthermore, we find that unlike granulation, supergranulation can produce peaks in periodograms that are significantly above the false-alarm probability level, even at a long period such as 300 d. When RV time series are analyzed using periodograms and fap, great care is therefore required. Subsets of the data show peaks at different periods, however (the peaks are naturally not stable), which can be a criterion to distinguish between supergranulation and a planetary signal, but a very good temporal sampling is therefore required to allow subseries to be analyzed. 

Supergranulation could therefore be the main obstacle for solar-type stars when a 1 M$_{\rm earth}$  planet in the habitable zone around a G2 star is to be detected. Its signal cannot be easily averaged \cite[][]{meunier15}. It does not have any counterpart in photometry either \cite[the contrast is lower than 1K, e.g.,][]{meunier07c}, which excludes using this observable to help take this contribution to RV into account. It is also unlikely that supergranulation will produce line distortions that can be identified and used because they are probably much smaller than those due to activity and granulation. 
In future work, correction techniques using Gaussian processes will be tested. Periodogram standardizations such as proposed by \cite{sulis16,sulis17} for granulation could also be tested.

The mass characterization is also affected by this type of signal. For only a few points, that is, a number corresponding to detection rates close to zero, the 1$\sigma$ uncertainty on the mass is between 39\% and 62 \% depending on the considered supergranulation level. It falls to 23\%-45\% in a regime with a good detection rate (1500 points).

We finally recall that this source of variable RV in the measurement is added to other sources that in particular are due to magnetic activity, such as spots and plages \cite[e.g.,][]{lagrange10b,meunier10a}, which also hampers planet detection: \cite{borgniet15} obtained detection limits close to 2 M$_{\rm earth}$ for a planet in the habitable zone of a G2 star when they assumed a solar activity level. 
We conclude that with the current knowledge, finding Earth twins around Sun-like stars may be beyond our current capabilities if these amplitudes are confirmed. Concerning granulation, the simulations made in \cite{meunier15} were based on granular flows that were very realistic because this type of hydrodynamical simulations reproduces the line profiles very well \cite[][]{asplund00}. The center-to-limb variations were well taken into account, although some fine effects in line properties across the disk \cite[e.g.,][]{asplund00, cegla18} were simplified, which may result in a slightly different rms RV due to granulation. For supergranulation, the uncertainties were included in the estimations made in \cite{meunier15} because they provided upper and lower limits: this was taken into account in the present paper as well. We also note that the stellar amplitudes in \cite{dumusque11b} were obtained from fits of the power spectra, but these the uncertainties in these estimates are not well characterized.

In a future paper, we will investigate in more detail how the power due to supergranulation affects the exoplanet detectability for other spectral types and a wider range of orbital periods, and whether a more suitable temporal sampling could help mitigate this contribution in a better way.

\begin{acknowledgements}

This work has been funded by the ANR GIPSE ANR-14-CE33-0018.
This work was supported by the "Programme National de Physique Stellaire" (PNPS) of CNRS/INSU co-funded by CEA and CNES and by the Programme National de Plan\'etologie (PNP) of CNRS/INSU, co-funded by CNES.
We thank the anonymous referee for his/her useful comments which helped to improve the paper.

\end{acknowledgements}

\bibliographystyle{aa}
\bibliography{biblio}

%

\end{document}